\documentstyle[preprint,aps]{revtex}
\begin{document}
\draft
\title{\bf {Topological effects on magnitude of persistent
current in a ring }}
\author{P. Singha Deo\cite{eml} }
\address{Institute of Physics, Bhubaneswar 751005, India}
\maketitle
\begin{abstract}
We show that defects in 1D rings
decrease the conductance (when the ring is opened up)
many more times than it decreases the
persistent currents. This means that the states in such a 1D
ring are very sensitive to twisting of boundary condition but
conductance of the system is small.
In 1D the electron
effectively sees a periodic potential and escapes localization.
This does not happen in higher dimensions. This helps us to
understand the simulations of G. Kirczenow[13] and also suggests
that a rough boundary in a 3D ring
that provide effectively a strong 1D potential to the electron
will have this type of different effects on conductance and
persistent currents.
\end{abstract}
\narrowtext
\newpage

Much before the experimental observations B$\ddot u$ttiker et
al [1] had predicted that equilibrium persistent currents may be
observed in a normal metal or semiconductor ring. But the
experimental observations[2-4] of these currents have posed many
serious problems. Prominent among these problems is that
persistent currents observed in a single gold ring[3] is
orders of magnitude larger than that predicted
theoretically.

There has been a large number of works[5-7] trying to give a many
body explanation for the large observed value of persistent
current. But recent detailed calculations[8-11] show
no significant enhancement of persistent currents. Now it is
considered that search for a many body explanation of the
experiments of Chandrasekhar et.  al. is far from over[12].

In the experiment[3] the conductance of the ring was measured by
opening up the ring and forming it into a wire. The conductance
across it was measured and the elastic mean free path was
estimated from that. Then it was estimated that the small
elastic mean free path of
electron thus obtained cannot sustain the large value of
observed persistent current. Ref [13] has shown that a special kind of
defects known as grain boundaries can decrease the conductance
much more drastically than the persistent currents in a 3D
multichannel ring and this does
not happen with point defects. Such defects as grain boundaries
are possible
in a gold ring and in case such defects are  present (can
be observed with a microscope) then it is incorrect to estimate
the persistent current from the conductance.
In this paper we show that it is a basic feature of 1D rings to
exhibit this effect where an impurity can decreases conductance many
more times than it decreases persistent currents. This helps us
to bring out deeper physics guiding the behavior of an electron
in a ring. Understanding the basic physics help us to argue that
another type of defect that
are invariably present in the experimental rings, can
produce similar effects.

To this end  we consider a single delta potential impurity
in a ring of length L and consider the electrons in the ring to
be free. To measure the conductance of the ring in
the sense of Chandrasekhar et. al. we have to severe the ring at
a point say P shown in fig 1 and form it into a wire shown in
fig 2. The two
ends of the wire i.e. A and B are then connected to two
semi-infinite leads to calculate the conductance using
Landauer's conductance formula. The persistent current is
calculated using the procedure of ref [14] based on first
principles that involves no
approximations. After some simplification we find that for our
system the allowed modes k are given by the solution of the
following equation.

\begin{equation}
cos(\alpha)={Vsin(kL) \over 2k} + cos(kL)
\end{equation}

\noindent where $\alpha$ is the Ahoronov Bohm phase aquired by the
electron in moving round the ring once and V is the strength of
the delta potential in the ring. The persistent current at a
certain allowed value of k is given by

\begin{equation}
I_k=(-e/h){2ksin(\alpha) \over {VLcos(kL) \over 2k} -sin(kL)({V \over
2k^2} +L)}
\end{equation}

\noindent Table 1. and table 2. lists a few values showing by how
much the impurity decreases the conductance and by how much it
decreases the total persistent current. The value of parameters
chosen is given in the table captions.

As we are studying the relative change in conductance and
persistent current we can calculate the persistent current for
any value of $\alpha$ but we have calculated it at a value of
$\alpha=\pi/2$ because the magnitude of the persistent current
at this value is approximately maximum.  I(n) means the total
persistent current when n electrons are present in the ring with
a delta potential impurity of strength V. $I_0(n)$ is the total
persistent current when n electrons are present in a clean ring.
Fermi energy of a ring with a delta potential and having n
electrons is slightly higher than that of a clean ring with n
electrons. $g_E$ is the conductance evaluated at a Fermi energy
(chemical potential of the reservoir determines this Fermi
energy) which corresponds to the Fermi energy of the ring with
the delta potential (the one that has a higher Fermi energy;
although we could have well evaluated at a Fermi energy which is
average of the Fermi energies of the two rings).
$g^0_E$ is the conductance of a
clean wire which at all Fermi energy is $2e^2/h$. For a few
electrons in the ring it is found that the impurity decreases
the conductance many more times than it decreases the persistent
currents(see table 1). This effect becomes smaller as the number
of electrons in the ring increases but the conductance always
decreases more than the persistent currents.  As more electrons
are put in the ring the higher energy states get filled up. It
is well known that higher energy electrons do not feel the delta
potential and if the potential is not felt at all it cannot be
expected to produce different types of effect on persistent
currents and conductance. We can increase the length of the ring
and thus decrease the level spacings to create more levels in
the region where the potential is strongly felt. Note that
this does not affect the conductance of the ring
(when opened up) because in the scattering problem this
length scale is not
involved in any way as will be explained soon.
Table 2 compares the decrease in persistent current
and conductance in this case. We find that to a larger filling
the difference between decrease in persistent current and
decrease in conductance is formidable. It also shows that for
the first 2
levels this difference has increased by another order of
magnitude.  This formidable difference (orders of magnitude
difference) when the delta potential is strongly felt is very
counterintutive. It is well known that more is the sensitivity
of the states to twisting of boundary condition more will be the
persistent current carried by the states. It is also well known
that more is the sensitivity of the states to twisting of boundary
condition more will be the  conductance of the system.
But here we have a simplest
situation where the states are very sensitive to twisting of
boundary condition and yet conductance is very small.

The mechanism by which the impurity can decrease persistent
current is different from that by which it can decrease the
conductance and this can make a lot of difference in 1D.
When we are finding the conductance we are just solving the
scattering problem of a geometry shown in fig 2. There is
absolutely no length scale in the problem as is evidient from
fig. 2. But as soon as we
join the points A and B to form a ring (for the time being there
is no magnetic field) there is a length scale
which is the length L of the ring. Ref [15] has shown
that even in absence of magetic field the electron in a 1D ring
is moving  with a momentum K given by KL=re(1/t) which is
exactly the Block momentum of a periodic system of delta
potentials of strength V and L is the length
separating the delta potentials. This means the electron is
effectively moving in such a periodic potential system.
This does not happen in higher dimensions. Now if we put a
magnetic field the
effect of the magnetic field is to twist the boundary
condition and produce a dispersion with $\phi$. This $\phi$
is called a
pseudo Block momentum[1]. So as soon as we connect points A and B
we change the topology of the system in which the electron is
moving. As a result in its motion the electron encounters an
identical delta potential after traversing identical path
lengths whereas earlier in case of the scattering problem it
could encounter the delta potential
only once and then get dissipated into the reservoir. It is well
known that the probability of scattering
by an infinite periodic array of potentials in certain range of
Fermi energies (band energies)
is much too smaller than the probability of
scattering by any one of the delta potentials making the
periodic array and this difference increases as the strength
of the delta potential increases. But as the strength increases
this range of Fermi energy or the band becomes narrow.
At other range of Fermi energies (band gap energies) where the
probability of scattering by an infinite periodic potential is
much smaller than the probability of scattering by single delta
potential, most probably we do not get an allowed mode in the
ring. This is so because these modes are evanescent modes and
cannot carry current. Only under some special
situations one can excite evanescent modes that contribute to persistent
currents[15,16]. So an electron in any
one of the allowed modes in the ring experience very little
resistance from the scatterer than it experiences when the ring
is cut open and the electron is incident on the scatterer from
one side.

When we
are studying the conductance the impurity breaks translational
symmetry completely and hence scatters vigorously. But as soon
as we form it into a ring (we have not put any magnetic field
yet) a discrete symmetry is restored. As
now it is a ring it has a rotational symmetry instead of a
translational symmetry and it corresponds to rotations like
2$\pi$, 4$\pi$, 6$\pi$ etc. Hence we have symmetry dictated good
quantum numbers in the ring (1D ring) called topological quantum
numbers and they never mix. An electron left to a state
characterized by a topological quantum number will stay in this
state for ever unscattered and it will have a definite group
velocity constant in dirrection and magnitude. It is just like
the case of an
electron that can pass through a discrete but regular lattice
without being scattered.  However increasing the potential
strength narrows the band and decrease the group velocity of the
electron. It is this decrease in group velocity that decreases
the persistent current and not the scattering.
In higher dimensions also we have the same
discrete rotational symmetry and topological quantum numbers
along with a transverse or subband quantum number. There are
actually a few propagating subbands that are degenerate and
impurities
can mix these different subbands and hence the different
topological states having the same total energy. This is in spite
of the discrete rotational symmetry.
However due to
the transverse degree of freedom an additional effect comes into
play which does not happen in 1D rings. The second time the
electron goes round the ring it may be encountering a different
disorder configuration than it did the first time because of the
transverse degree of freedom. So in higher dimension the
electron in a particular state charecterized by a topological
quantum number will not see an effective periodic potential but
rather an effective
random potential in the absence of magnetic field. Hence it has
an extremely small group velocity. Smaller is the group velocity
lesser will be the sensitivity of the state to twisting of
boundary condition by the magnetic field and smaller will be the
persistent current carried by it. Hence the amount by which the impurities
decrease the conductance is comparable to the amount by which it
decreases the sensitivity of the states to twisting of boundary
conditions. This is what is found in the simulations of ref[13] that
for point defects in a ring the persistent current decreases as many
times as the conductance.
But the physics is completely different in 1D. When the ring is closed
the electrons feel a periodic potential and are very sensitive to
twisting of boundary condition. Ref[15] has shown that for a
ring with defects the equation that determines the twisting of
boundary condition with magnetic field i.e. $\alpha$ is
$e^{i(KL+\alpha)}$ where K is the momentum in the periodic
system and hence the states can be very sensitive to the
twisting of boundary condition. As soon as the ring is opened up
the electrons do not feel the periodic potential and hence contribute very
little to conduction. The numbers given in the table and their
orders of magnitude establish this.

Grain boundaries are extended defects that run across the whole
cross section of the ring. If the grain boundaries are radial
then they are more effective in decreasing the conductance much
more than the persistent current. For the time being consider a
single radial grain boundary.  If the defect be such as the
grain boundary then again in higher dimensional ring also the
electron will encounter the grain boundary at regular intervals
in going round the ring again and again.
This also explains why non radial grain
boundaries are worse than radial ones but obviously they will be
beter than point defects.

It is also well known that boundary roughness can be mapped into
an effective 1D potential[17] and it has also been shown that
smaller is the boundary roughness greater will be the strength
of the effective 1D potential[15]. Because of the later fact
it is
quiet possible that there are very strong effectively 1D
potentials in the experimental ring and hence they will exhibit
the same effect of decreasing the conductance much more than
decreasing the persistent currents. A simulation of boundary
roughness in a multichannel ring will be reported very soon.

\vfill
\eject

\vfill
\eject
\end{document}